\documentclass[12pt]{article}

\usepackage{natbib}
\usepackage[nottoc,notlof,notlot]{tocbibind} 

\bibpunct{(}{)}{,}{a}{}{;}
\usepackage{graphicx}
\graphicspath{ {images/} }

\usepackage{amsmath}
\usepackage{bbm}
\usepackage{amsfonts}%
\usepackage{amssymb}%

\usepackage{geometry} 
\usepackage{graphicx} 
\usepackage{framed}
\usepackage{caption}
\usepackage{subcaption}

\usepackage{booktabs} 
\usepackage{array} 
\usepackage{paralist} 
\usepackage{verbatim} 

\usepackage{lipsum}
\usepackage{setspace}

\usepackage{lipsum}
\usepackage[singlelinecheck=false]{caption}
\usepackage{tabularx}

\usepackage[colorlinks]{hyperref}
\hypersetup{allcolors=blue}
\title{\setstretch{1}Identification and Estimation of a Partially Linear Regression Model using Network Data}

\author{Eric Auerbach\footnote{Department of Economics, Northwestern University. E-mail: eric.auerbach@northwestern.edu. I thank my advisors, James Powell and Bryan Graham for their advice and support. I also thank Jonathan Auerbach, Ivan Canay, David Card, Christina Chung, Aluma Dembo, Joel Horowitz, Michael Jansson, Patrick Kline, Sheisha Kulkarni, Chuck Manski, Konrad Menzel, Carl Nadler, Stephen Nei, Aureo de Paula, Demian Pouzo, Mikkel Soelvsten, Katalin Springel, Max Tabord-Meehan and participants at the UC Berkeley Econometrics Seminar for helpful feedback.}} \date{\parbox{\linewidth}{\centering%
  \today\endgraf}} 

\begin{document}
\maketitle
\begin{abstract} \setstretch{1}\noindent
I study a regression model in which one covariate is an unknown function of a latent driver of link formation in a network. Rather than specify and fit a parametric network formation model, I introduce a new method based on matching pairs of agents with similar columns of the squared adjacency matrix, the $ij$th entry of which contains the number of other agents linked to both agents $i$ and $j$. The intuition behind this approach is that for a large class of network formation models the columns of the squared adjacency matrix characterize all of the identifiable information about individual linking behavior. In this paper, I describe the model, formalize this intuition, and provide consistent estimators for the parameters of the regression model. \cite{auerbach2021identification} considers inference and an application to network peer effects. 
\end{abstract}

\section{Introduction}
Most economic outcomes are not determined in isolation. Rather agents are influenced by the behaviors and characteristics of other agents. For example, a high school student's academic performance might depend on the attitudes and expectations of that student's friends and family \citep[see generally][]{sacerdote2011peer,bramoulle2019peer}. 

Incorporating this social influence into the right-hand side of an economic model may be desirable when the researcher wants to understand its impact on the outcome of interest or when it confounds the impact of another explanatory variable such as the causal effect of some nonrandomized treatment.  For instance, the researcher may want to learn the causal effect of a tutoring program on academic performance in which program participation and counterfactual academic performance are both partially determined by family expectations. However, in many cases the relevant social influence is not observed by the researcher. That is, the researcher does not have access to data on the family expectations that confound the causal effect of the tutoring program and thus cannot control for this variable using conventional methods. 

An increasingly popular solution to this problem is to collect network data and suppose that the unobserved social influence is revealed by linking behavior in the network. For instance, the researcher might observe pairs of students who identify as friends and believe that students with similar reported friendships have similar family expectations. It is not immediately clear, however, how one might actually use network data to account for this unobserved social influence in practice, since the number of ways in which agents can be linked in a network is typically large relative to the sample size. 

This paper proposes a new way to incorporate network data into an econometric model. I specify a joint regression and network formation model, establish sufficient conditions for the parameters of the regression model to be identified, and provide consistent estimators. Large sample approximations for inference and an application to network peer effects building on work by \cite{bramoulle2009identification, de2010identification, goldsmith2013social, hsieh2014social, johnsson2015estimation,arduini2015parametric}, and others is provided by \cite{auerbach2021identification}. 

A limitation of the framework is that the large sample approximations suppose a sequence of networks that is asymptotically dense in that the fraction of linked agent-pairs does not vanish with the sample size. The regime can fail to characterize potentially relevant features of networks in which relatively few agent pairs are linked \cite[see][for a discussion]{mele2017structuralb}. Potential extensions to sparse asymptotic regimes are left to future work.

\section{Framework}

\subsection{Model}
Let $i$ represent an arbitrary agent from a large population. Associated with agent $i$ are an outcome $y_{i} \in \mathbb{R}$, an observed vector of explanatory variables $x_{i} \in \mathbb{R}^{k}$, and an unobserved index of social characteristics $w_{i} \in [0,1]$. The three are related by the model
\begin{align}
y_{i} &= x_{i}\beta + \lambda(w_{i}) + \varepsilon_{i} \label{eq2} 
\end{align}
where $\beta \in \mathbb{R}^{k}$ is an unknown slope parameter, $\lambda$ is an unknown measurable function, and  $\varepsilon_{i}$ is an idiosyncratic error. 

The researcher draws a sample of $n$ agents uniformly at random from the population. This sample is described by the sequence of independent and identically distributed random variables $\{y_{i},x_{i}, w_{i}\}_{i=1}^{n}$, although only $\{y_{i},x_{i}\}_{i=1}^{n}$ is observed as data. The researcher also observes $D$, an $n\times n$ stochastic binary adjacency matrix corresponding to an unlabeled, unweighted, and undirected random network between the $n$ agents. The existence of a link between agents $i$ and $j$ is determined by the model
\begin{align}
D_{ij} &=   \mathbbm{1}\{\eta_{ij} \leq f(w_{i},w_{j})\}\mathbbm{1}\{i \neq j\}  \label{eq3}
\end{align}
in which $f$ is a symmetric measurable function satisfying the continuity condition that $\inf_{u \in [0,1]}\int \mathbbm{1}\left\{v \in [0,1]: \sup_{\tau \in [0,1]}\left|f(u,\tau)-f(v,\tau)\right| < \varepsilon\right\}dv > 0$ for every $\varepsilon > 0$ and $\{\eta_{ij}\}_{i,j=1}^{n}$ is a symmetric matrix of unobserved scalar disturbances with independent upper diagonal entries that are mutually independent of $\{x_{i},w_{i},\varepsilon_{i}\}_{i=1}^{n}$. This continuity condition is weaker than the typical assumption that $f$ is a continuous function. It is used because it allows for a variety of models where $f$ is ``almost'' but not quite a continuous function. For example, in the blockmodel described in Section 2.2.1 below, $f$ is  a piecewise continuous function. 


The regression model (\ref{eq2}) represents a pared-down version of various linear models popular in the network economics literature. For instance in the network peer effects literature, $y_{i}$ could be student $i$'s GPA, $x_{i}$ could indicate whether $i$ participates in a tutoring program, $w_{i}$ could index student $i$'s participation in various social cliques, and $\lambda(w_{i})$ could represent the influence of student $i$'s peers' expected GPA, program participation, or other characteristics on student $i$'s GPA. That is, supposing $P(D_{ij} = 1 |w_{i}) > 0$,
\begin{align*}
\lambda(w_{i}) = E\left[x_{j}|D_{ij}=1,w_{i}\right]\gamma + \delta E\left[y_{j}|D_{ij}=1,w_{i}\right]
\end{align*}
for some $(\gamma,\delta) \in \mathbb{R}^{k+1}$ \citep[see relatedly][]{manski1993identification}. To demonstrate the proposed methodology, this paper conflates these different possible social effects into one social influence term, $\lambda(w_{i})$. This may be sufficient to identify the impact of the tutoring program holding social influence constant, predict a student's GPA under some counterfactual social influence, or test for the existence of any social influence. \cite{auerbach2021identification} discusses how one can also separately identify different social effects.

The parameters of interest are $\beta$ and $\lambda(w_{i})$, the realized social influence for agent $i$. The function $\lambda: [0,1] \to \mathbb{R}$ is not a parameter of interest because it is not separately identified from $w_{i}$ (see Section 2.2.1 below). It is without loss to normalize the distribution of $w_{i}$ to be standard uniform.

Network formation (\ref{eq3}) is represented by $n \choose 2$ conditionally independent Bernoulli trials. The model is a nonparametric version of a class of dyadic regression models popular in the network formation literature. Section 6 of \cite{graham2019network} or Section 3 of \cite{de2020econometric} contains many examples. It is often given a discrete choice interpretation in which $f(w_{i},w_{j}) - \eta_{ij}$ represents the marginal transferable utility agents $i$ and $j$ receive from forming a link, which precludes strategic interactions between agents. The distribution of $\eta_{ij}$ is not separately identified from $f$ and so is also normalized to be standard uniform. 

Under  (\ref{eq3}), the observed network $D$ is almost surely dense or empty in the limit. That is, for a fixed $f$ and as $n$ tends to infinity, ${n \choose 2}^{-1}\sum_{i=1}^{n-1}\sum_{j=i+1}^{n}D_{ij}$ will either be bounded away from zero or exactly zero with probability approaching one. The framework can potentially accommodate network sparsity by allowing $f$ to vary with the sample size (see Appendix Section A.1), but a formal study of such an asymptotic regime is left to future work. 

The following Assumption 1 collects key aspects of the model for reference. 
\newline

\begin{flushleft}
\textbf{Assumption 1}: The random sequence $\{x_{i},w_{i},\varepsilon_{i}\}_{i=1}^{n}$ is independent and identically distributed with entries mutually independent of $\{\eta_{ij}\}_{i,j = 1}^{n}$, a symmetric random matrix with independent and identically distributed entries above the diagonal. The outcomes $\{y_{i}\}_{i=1}^{n}$ and $D$ are given by equations (\ref{eq2}) and (\ref{eq3}) respectively. The variables $x_{i}$ and $\varepsilon_{i}$ have finite eighth moments, $w_{i}$ and $\eta_{ij}$ have standard uniform marginals, $E\left[\varepsilon_{i}|x_{i}, w_{i}\right] = 0$, $\inf_{u \in [0,1]}\int \mathbbm{1}\left\{v \in [0,1]: \sup_{\tau \in [0,1]}\left|f(u,\tau)-f(v,\tau)\right| < \varepsilon\right\}dv > 0$ for every $\varepsilon > 0$, $\sup_{u \in [0,1]}\left|E\left[x_{i}|w_{i} = u\right]\right| < \infty$, $\sup_{u \in [0,1]}|\lambda(u)| < \infty$, and $0 \leq \inf_{u,v \in [0,1]}f(u,v) \leq$ $\sup_{u,v \in [0,1]}f(u,v) \leq 1$. 
\end{flushleft}

\subsection{Identification}
\subsubsection{Non-identification of the social characteristics}
If $w_{i}$ were observed, (\ref{eq2}) would correspond to the partially linear regression of \cite{robinson1988root} and the identification problem would be well-understood. If $w_{i}$ were unobserved but identified, one might replace $w_{i}$ with an empirical analogue as in \cite{ahn1993semiparametric}. Identification strategies along these lines are considered by \cite{arduini2015parametric} and \cite{johnsson2015estimation}. 

However, it is not generally possible to learn $w_{i}$ in the setting of this paper. The main difficulty is that many assignments of agents to social characteristics generate the same distribution of network links. Specifically, for any  measure-preserving invertible $\varphi$ (that is for any measurable $A \subseteq [0,1]$, $A$ and $\varphi^{-1}(A)$ have the same measure), $\left(\{w_{i}\}_{i=1}^{n},f(\cdot,\cdot)\right)$ and $\left(\{\varphi(w_{i})\}_{i=1}^{n},f(\varphi^{-1}(\cdot),\varphi^{-1}(\cdot))\right)$ generate the same distribution of links, where $w_{i}$ and $\varphi(w_{i})$ may be very different.  For example, if $\{w_{i}\}_{i=1}^{n}$ and $f(u,v) = (u+v)/2$ explain the distribution of $D$, then so too does $\{w_{i}'\}_{i=1}^{n}$ and $f'(u,v) = 1-(u+v)/2$ where $w_{i}' = 1-w_{i}$. 

Furthermore, even if the researcher is willing to posit a specific $f$, the social characteristics may still not be identified. For example, in a simplified version of the blockmodel of \cite{holland1983stochastic}, there exists an $l\times l$ dimensional matrix $\Theta$ such that $f(w_{i},w_{j}) = \Theta_{\lceil l w_{i}\rceil \lceil l w_{j}\rceil}.$ Intuitively, $[0,1]$ is divided into $l$ partitions (with agent $i$ assigned to partition $\lceil lw_{i}\rceil$) and the probability two agents link only depends on their partition assignments. In this case, the probability that agents link is invariant to changes in the social characteristics that do not change the agents' partition assignments, and so while the underlying partition assignments might be learned from $D$, the social characteristics that determine the partition assignments generally cannot. Notice that $f$ in this case is not a continuous function, but satisfies the continuity condition of Assumption 1.

Another example in which the social characteristics are not identified is the homophily model $f(w_{i},w_{j}) = 1-(w_{i}-w_{j})^{2}$. Intuitively, agents are more likely to form a link if their social characteristics are similar. In this case, both $\{w_{i}\}_{i=1}^{n}$ and $\{1-w_{i}\}_{i=1}^{n}$ generate the same distribution of links. 

An example in which the social characteristics are identified is the nonlinear additive model $f(w_{i},w_{j}) = \Lambda(w_{i}+w_{j}),$ where $\Lambda$ is a strictly monotonic function such as the logistic function \citep[see][Section 6.3]{graham2019network}. Intuitively, agents with larger values of $w_{i}$ are more likely to form links. In this case, $w_{i}$ is identified from $D$ because $w_{i} = P\left(\int \Lambda(w+\tau)d\tau \leq \int \Lambda(w_{i}+\tau)d\tau| w_{i}\right)$ where $\int \Lambda(w_{i}+\tau)d\tau = P\left(D_{it} = 1| w_{i}\right)$ and $w$ is an independent copy of $w_{i}$. 

\subsubsection{Agent link function}
Since $w_{i}$ is not generally identified, I propose an alternative description about how $i$ is linked in the network that is identified. I call this alternative an agent link function and propose using link functions instead of social characteristics to identify $\beta$ and $\lambda(w_{i})$. 

Agent $i$'s link function is the projection of $f$ onto $w_{i}$. That is, $f_{w_{i}}(\cdot) := f(w_{i},\cdot): [0,1] \to [0,1]$. It is the collection of probabilities that agent $i$ links to agents with each social characteristic in $[0,1]$. I consider link functions to be elements of $L^{2}([0,1])$, the usual inner product space of square integrable functions on the interval. I sometimes use $d(w_{i},w_{j}) := ||f_{w_i} - f_{w_j}||_{2}$ to refer to the pseudometric on $[0,1]$ induced by $L^2$-differences in link functions. I call this pseudometric network distance.  

Conditional expectations with respect to $f_{w_{i}}$ implicitly refer to the random variable $w_{i}$. For example, $E\left[x_{i} |\hspace{1mm} f_{w_{i}}\right]  := \lim_{h \to 0}E\left[x | \hspace{1mm} w \in \{u \in [0,1] : ||f_{u}-f_{w_{i}}||_{2} \leq h\},w_{i}\right]$ and \\$E\left[x_{i}x_{j}'  |\hspace{1mm} ||f_{w_{i}} - f_{w_{j}}||_{2} = 0\right]$ $:= \lim_{h\to0}E\left[x\tilde{x}' |\hspace{1mm}  (w,\tilde{w}) \in \{(u,v) \in [0,1]^{2} : ||f_{u}-f_{v}||_{2} \leq h\}\right]$ where $(x,w)$ and $(\tilde{x},\tilde{w})$ are independent copies of $(x_{i},w_{i})$. The conditional expectations on the right-hand side are well-defined for any $h > 0$ because of the continuity condition on $f$ in Assumption 1. Whenever the conditional expectation on the left-hand side is used, the relevant limit is assumed to exist.    


Under (\ref{eq3}), the link function $f_{w_{i}}$ is the totality of information that $D$ contains about $w_{i}$. It describes the law of the $i$th row of $D$ and so is identified. Furthermore, $w_{i}$ is only identified when $f_{w_{i}}$ is invertible in $w_{i}$. For example, in the nonlinear additive model  from Section 2.2.1, $w_{i}$ is identified because $u > v$ implies that $f_{u}(\cdot) := \Lambda(u + \cdot)$ dominates $f_{v}(\cdot) := \Lambda(v + \cdot)$. In this example, agents with different social characteristics necessarily have different probabilities of forming links to other agents in the population. In the blockmodel, $w_{i}$ is not identified because if $\lceil l u\rceil = \lceil l v\rceil$ then $f_{u}(\cdot) := \Theta_{\lceil l u\rceil \lceil l \cdot\rceil} =  f_{v}(\cdot) := \Theta_{\lceil l v\rceil \lceil l \cdot\rceil}$ even if $u \neq v$. In this example, agents with different social characteristics but the same partition assignment have the same probability of forming links to other agents in the population. 

The large-sample limits of many popular agent-level network statistics are determined by the agent's link function. Examples include degree $\frac{1}{n}\sum_{t=1}^{n}D_{it} \to_{p} E\left[D_{it}|w_{i}\right] = \int f_{w_{i}}(\tau)d\tau$, average peers' characteristics $\frac{\sum_{t=1}^{n}x_{t}D_{it}}{\sum_{t=1}^{n}D_{it}} \to_{p} E\left[x_{t}|D_{it} = 1,w_{i}\right] = \frac{\int E\left[x_{t}|w_{t} = \tau\right]f_{w_{i}}(\tau)d\tau}{\int f_{w_{i}}(\tau)d\tau}$, and clustering $\frac{\sum_{j=1}^{n-1}\sum_{k = j+1}^{n}D_{ij}D_{ik}D_{jk}}{\sum_{j=1}^{n-1}\sum_{k = j+1}^{n}D_{ij}D_{ik}} \to_{p} \frac{E\left[D_{ij}D_{ik}D_{jk}|w_{i}\right]}{E\left[D_{ij}D_{ik}|w_{i}\right]} = \frac{\int\int f_{w_{i}}(\tau)f_{w_{i}}(s)f(\tau,s)d\tau ds}{\left(\int f_{w_{i}}(\tau)d\tau\right)^{2}}$ (supposing $\int f_{w_{i}}(\tau)d\tau > 0$). This observation will partly inform Assumption 3 below.

\subsubsection{Identification of the regression model}
If $w_{i}$ were observed or identified, the standard approach would be to first identify $\beta$ using covariation between $y_{i}$ and $x_{i}$ unrelated to $w_{i}$ and then to identify  $\lambda(w_{i})$ using residual variation in $y_{i}$. This identification strategy requires variation in $x_{i}$ not explained by $w_{i}$. Let $\Xi(u) = E\left[(x_{i} - E\left[x_{i}|w_{i} \right])'(x_{i} - E\left[x_{i}|w_{i} \right])|w_{i} = u\right]$. 
\begin{flushleft}
\textbf{Assumption 2}: $\inf_{u \in [0,1]}\sigma_{k}(\Xi(u)) > 0$ where $\sigma_{k}(\cdot)$ is the smallest eigenvalue. 
\end{flushleft}

Assumption 2 is strong but standard. It is violated when the covariates include population analogues of network statistics such as agent degree or average peers' characteristics (or any other function of $w_{i}$). In such cases, alternative assumptions are required for identification. 

Since $w_{i}$ is neither observed nor identified, the standard approach cannot be implemented. A contribution of this paper is to propose using $f_{w_{i}}$ instead of $w_{i}$ for identification. The substitution relies on an additional assumption that $\lambda(w_{i})$ is determined by $f_{w_{i}}$.
\begin{flushleft}
\textbf{Assumption 3}: For every $\epsilon > 0$ there exists a $\delta > 0$ such that $\sup_{u,v \in [0,1]: ||f_{u}-f_{v}||_{2} \leq \delta}(\lambda(u)-\lambda(v))^{2} \leq \epsilon$.
\end{flushleft}

Assumption 3 is strong and new. In words, it says that agents with similar link functions have similar social influence. Since, under (\ref{eq3}), $f_{w_{i}}$ is the totality of information that $D$ contains about $w_{i}$, Assumption 3 supposes that this information is sufficient to discern $\lambda(w_{i})$. It does not restrict the function $f$.

One justification for the assumption could be that $w_{i}$ does not directly impact $y_{i}$. Instead, $w_{i}$ only influences $y_{i}$ by altering $i$'s linking behavior $f_{w_{i}}$. For example, if $w_{i}$ indexes student $i$'s participation in various social cliques, then the assumption follows if this index only directly affects which other students and teachers $i$ interacts with, and it is these interactions that ultimately determine $i$'s participation in the tutoring program and GPA. 

The assumption is also satisfied when the social influence is the population analogue of one of the network statistics described in Section 2.2.2. This is the case for the network peer effects example of Section 2.1 where $\lambda(w_{i}) = E\left[x_{j}|D_{ij}=1,w_{i}\right]\gamma + \delta E\left[y_{j}|D_{ij}=1,w_{i}\right]$, because $E\left[z_{j}|D_{ij}=1,w_{i}\right] = \frac{\int E\left[z_{j} | w_{j} = \tau\right]f_{w_{i}}(\tau)d\tau}{\int f_{w_{i}}(\tau)d\tau}$ is a continuous functional of $f_{w_{i}}$.


However, the assumption may be implausible when the network is sparse (the link function is close to $0$) because every agent-pair may have network distance close to zero. As a result, under network sparsity, Assumption 3 may imply that $\lambda(w_{i})$ behaves like a constant. See Appendix Section A.1 for a discussion.



Proposition 1 states that Assumptions 1-3 are sufficient for $\beta$ and $\lambda(w_{i})$ to be identified.
\begin{flushleft}
\textbf{Proposition 1}: Under Assumptions 1-3,
\begin{itemize}
\item[(i)]  $\beta = \text{argmin}_{b \in \mathbb{R}^{k}} E\left[\left(y_{i}-y_{j} - (x_{i}-x_{j})b\right)^{2}|\hspace{1mm}||f_{w_i}-f_{w_j}||_{2} = 0\right]$ and
\item[(ii)] $\lambda(w_{i}) = E\left[\left(y_{i} - x_{i}\beta\right) |\hspace{1mm}f_{w_{i}}\right]$.
\end{itemize}
\end{flushleft} 


I close with two examples in which $\beta$ and $\lambda(w_{i})$ are identified (Assumptions 1-3 hold) but  $w_{i}$ is not. The first example is the case where links are determined by a blockmodel $f(w_{i},w_{j}) = \Theta_{\lceil lw_{i}\rceil\lceil lw_{j}\rceil}$ and social influence is determined by the agent partition assignments $y_{i} = x_{i}\beta + \alpha_{\lceil lw_{i}\rceil} + \varepsilon_{i}$. In this case, $\beta$ and $\alpha_{\lceil lw_{i}\rceil}$ are identified even though $w_{i}$ is not. The second example is the case where links are determined by a homophily model $f(w_{i},w_{j}) = 1-(w_{i}-w_{j})^{2}$ and social influence is an affine function of the agent social characteristics $y_{i} = x_{i}\beta + \rho_{1} +  \rho_{2} w_{i} + \varepsilon_{i}$. In this case $\beta$ and $\rho_{1} + \rho_{2} w_{i}$ are identified even though $(\rho_{1},\rho_{2})$ and $w_{i}$ are not separately identified.

\subsection{Estimation}
Estimation of $\beta$ and $\lambda(w_{i})$ is complicated by the fact that $f_{w_{i}}$ is unobserved and difficult to approximate directly. A contribution of this paper is to demonstrate that estimation is still possible using columns of the squared adjacency matrix. To explain the procedure, I introduce the codegree function. 

\subsubsection{Agent codegree function}
Let $p$ map $(w_i,w_j)$ to the conditional probability that $i$ and $j$ have a link in common, i.e. $p(w_{i},w_{j}) := \int f_{w_{i}}(\tau)f_{w_{j}}(\tau)d\tau$.  Agent $i$'s codegree function is  the projection of $p$ onto $w_{i}$. That is, $p_{w_{i}}(\cdot) := p(w_{i},\cdot): [0,1] \to [0,1]$. Codegree functions are also taken to be elements of $L^{2}([0,1])$. I sometimes use $\delta$ to refer to the  pseudometric on $[0,1]$ induced by $L^{2}$-differences in codegree functions,
\begin{align*}
\delta(w_{i},w_{j}) &:= ||p_{w_{i}} - p_{w_{j}}||_{2} = \left(\int \left(\int f(\tau,s)\left(f(w_{i},s)-f(w_{j},s)\right)ds\right)^{2}d\tau\right)^{1/2}.
\end{align*}
I call this pseudometric codegree distance. Conditional expectations with respect to codegree functions are defined exactly as they are for link functions.

In contrast to link functions, the population analogues of most network statistics (including those in Section 2.2.2) cannot naturally be written as functionals of codegree functions. The use of codegree functions is instead motivated by Lemma 1 below.

\subsubsection{Estimators}
I propose using codegree functions instead of link functions to construct estimators for $\beta$ and $\lambda(w_{i})$. The proposal relies on two results. The first result is that agents with similar codegree functions have similar link functions. The second result is that codegree distance can be consistently estimated using the columns of the squared adjacency matrix. 

The first result is given by Lemma 1 and is related to arguments from the link prediction literature \citep[see in particular][]{lovasz2010regularity,rohe2011spectral,zhang2015estimating}. 

\begin{flushleft} 
\textbf{Lemma 1}: If $0 \leq \inf_{u,v \in [0,1]}f(u,v) \leq \sup_{u,v \in [0,1]}f(u,v) \leq 1$ then for every $i, j \in \{1,...,n\}$ 
\begin{align*}
||p_{w_{i}} - p_{w_{j}}||_{2} \leq ||f_{w_{i}} - f_{w_{j}}||_{2}.
\end{align*}
If also $\inf_{u \in [0,1]}\int \mathbbm{1}\left\{v \in [0,1]: \sup_{\tau \in [0,1]}\left|f(u,\tau)-f(v,\tau)\right| < \epsilon\right\}dv > 0$ for every $\epsilon > 0$ then for every $i, j \in \{1,...,n\}$ and $\varepsilon >0$ there exists a $\delta > 0$ such that 
\begin{align*}
||f_{w_{i}} - f_{w_{j}}||_{2}\times \mathbbm{1}\{||p_{w_{i}} - p_{w_{j}}||_{2} \leq \delta\} \leq \varepsilon.
\end{align*}
\end{flushleft}

I defer a discussion of Lemma 1 to Section 2.3.3 and emphasize here instead its implication that the parameters of interest can be expressed as functionals of the agent codegree functions. That is, under Assumptions 1-3, $\beta$ uniquely minimizes \\$E\left[\left(y_{i}-y_{j} - (x_{i}-x_{j})b\right)^{2}|\hspace{1mm}||p_{w_i}-p_{w_j}||_{2} = 0\right]$ over $b \in \mathbb{R}^{k}$ and $\lambda(w_{i}) = E\left[\left(y_{i} - x_{i}\beta\right) |\hspace{1mm}p_{w_{i}}\right]$. 

The second result is that $\delta(w_{i},w_{j})$ can be consistently estimated by the root average squared difference in the $i$th and $j$th columns of the squared adjacency matrix,
\begin{align}\label{pd}
\hat{\delta}_{ij} :=  \left(\frac{1}{n}\sum_{t=1}^{n}\left(\frac{1}{n}\sum_{s=1}^{n}D_{ts}(D_{is}-D_{js})\right)^{2}\right)^{1/2}.
\end{align}

Intuitively, the empirical codegree $\frac{1}{n}\sum_{s=1}^{n}D_{ts}D_{is}$ counts the fraction of agents that are linked to both agents $i$ and $t$, $\{\frac{1}{n}\sum_{s=1}^{n}D_{ts}D_{is}\}_{t=1}^{n}$ is the collection of empirical codegrees between agent $i$ and the other agents in the sample, and $\hat{\delta}_{ij}$ gives the root average squared difference in $i$'s and $j$'s collection of empirical codegrees. That $\hat{\delta}_{ij}$ converges uniformly to $||p_{w_{i}} - p_{w_{j}}||_{2}$ over the $n\choose 2$ distinct pairs of agents as $n\to \infty$ is shown in Appendix Section A.4 as Lemma B1.


A consequence of these two results and Assumptions 1-3 is that when the $i$th and $j$th columns of the squared adjacency matrix are similar then $(y_{i}-y_{j})$ and $(x_{i}-x_{j})\beta + (\varepsilon_{i}-\varepsilon_{j})$ are approximately equal. Under additional regularity conditions provided in Section 2.3.4, $\beta$ is consistently estimated by the pairwise difference estimator
 \begin{align}\label{estimator}
\hat{\beta} = &\left( \sum_{i=1}^{n-1}\sum_{j = i+1}^{n}(x_{i}-x_{j})'(x_{i}-x_{j})K\left(\frac{\hat{\delta}^{2}_{ij}}{h_{n}}\right)\right)^{-1} \left(\sum_{i=1}^{n-1}\sum_{j = i+1}^{n}(x_{i}-x_{j})'(y_{i}-y_{j})K\left(\frac{\hat{\delta}^{2}_{ij}}{h_{n}}\right)\right)
\end{align}
and $\lambda(w_{i})$ is consistently estimated by the Nadaraya-Watson-type estimator
\begin{align}\label{otherestimator}
\widehat{\lambda(w_{i})} = \left(\sum_{t=1}^{n}K\left(\frac{\hat{\delta}^{2}_{it}}{h_{n}}\right)\right)^{-1}\left(\sum_{t=1}^{n}\left(y_{t} - x_{t}\hat{\beta}\right)K\left(\frac{\hat{\delta}^{2}_{it}}{h_{n}}\right)\right)
\end{align}
where $K$ is a kernel function and $h_{n}$ is a bandwidth parameter depending on the sample size. Since codegree functions are not finite-dimensional, the regularity conditions I provide for consistency are different than what is typically assumed. Conditions sufficient for the estimators to be asymptotically normal, consistent estimators for their variances, and more are provided by \cite{auerbach2021identification}.

\subsubsection{Discussion of Lemma 1}
The proof of Lemma 1 can be found in Appendix Section A.2. The first part, that $||p_{w_{i}} - p_{w_{j}}||_{2} \leq ||f_{w_{i}} - f_{w_{j}}||_{2}$, is almost an immediate consequence of Jensen's inequality. The second part is related to Theorem 13.27 of \cite*{lovasz2012large}, the logic of which demonstrates that $||p_{w_{i}} - p_{w_{j}}||_{2} = 0$ implies $||f_{w_{i}} - f_{w_{j}}||_{2} = 0$ when $f$ is a continuous function. The proof is short.
\begin{align*}
&||p_{w_{i}} - p_{w_{j}}||_{2}^{2} = 0 \implies \int\left(\int f(\tau,s)\left(f(w_{i},s)-f(w_{j},s)\right)ds\right)^{2}d\tau = 0 \\
&\implies \int f(\tau,s)\left(f(w_{i},s)-f(w_{j},s)\right)ds = 0 \text{ for every } \tau \in [0,1] \\
&\implies \int f(w_{i},s)\left(f(w_{i},s)-f(w_{j},s)\right)ds = 0 \text{ and } \int f(w_{j},s)\left(f(w_{i},s)-f(w_{j},s)\right)ds = 0 \\
&\implies \int \left(f(w_{i},s)-f(w_{j},s)\right)^{2}ds = 0 \implies ||f_{w_{i}} - f_{w_{j}}||_{2}^{2} = 0.
\end{align*}

Intuitively, if agents $i$ and $j$ have identical codegree functions then the difference in their link functions $(f_{w_{i}} - f_{w_{j}})$ must be uncorrelated with every other link function in the population, as indexed by $\tau$. In particular, the difference is uncorrelated with $f_{w_{i}}$ and $f_{w_{j}}$, the link functions of agents $i$ and $j$. However, this is only the case when $f_{w_{i}}$ and $f_{w_{j}}$ are perfectly correlated. 

Lov\'asz's theorem demonstrates that agent-pairs with identical codegree functions have identical link functions. The estimation strategy proposed in this paper, however, requires a stronger result that agent-pairs with similar but not necessarily identical codegree functions have similar link functions. This is the statement of Lemma 1. 

\cite{auerbach2021identification} derives rates of convergence for the estimators under a stronger version of Lemma 1. I include the result here as it may be of independent interest.
\begin{flushleft}
\textbf{Lemma A1}: Suppose $f$ satisfies $0 \leq \inf_{u,v \in [0,1]}f(u,v) \leq \sup_{u,v \in [0,1]}f(u,v) \leq 1$ and the $\alpha$-H\"older-continuity condition that there exists $\alpha, C > 0$ such that $\inf_{u \in [0,1]}\int \mathbbm{1}\left\{ v \in [0,1]: \sup_{\tau \in [0,1]}\left|f(u,\tau)-f(v,\tau)\right| < \varepsilon\right\} dv \geq \left(\frac{\varepsilon}{C}\right)^{1/\alpha}$  for every $\varepsilon \in [0,1]$. Then for every $i, j \in \{1,...,n\}$
\begin{align*}
||p_{w_{i}} - p_{w_{j}}||_{2} \leq ||f_{w_{i}} - f_{w_{j}}||_{2} \leq 2C^{\frac{1}{2+4\alpha}}||p_{w_{i}} - p_{w_{j}}||_{2}^{\frac{\alpha}{1+2\alpha}}. 
\end{align*}
\end{flushleft}

Lemma A1 bounds the cost of using codegree distance as a substitute for network distance in the estimation of $\beta$ and $\lambda(w_{i})$. Its proof can be found in Appendix Section A.2. When $C = \alpha = 1$, the result requires an agent-pair to have a codegree distance less than $\varepsilon^{3}/8$ to guarantee that their network distance is less than $\varepsilon$. The rate of convergence of the estimators based on codegree distance may be slower than the infeasible estimators based on network distance. 

\subsubsection{Consistency}
The following regularity conditions are imposed. Let $r_{n}(u) := \int K\left(\frac{||p_{u}-p_{v}||^{2}_{2}}{h_{n}}\right)dv$. 
\begin{flushleft}
\textbf{Assumption 4:} $h_{n} \to 0$, $n^{1-\gamma}h_{n}^{2} \to \infty$, and $\inf_{u \in [0,1]} n^{\gamma/4}r_{n}(u) \to \infty$ as $n \to \infty$ for some $\gamma > 0$. $K$ is nonnegative, twice continuously differentiable, and has support $[0,1)$. 
\end{flushleft}


The restrictions on $K$ are standard. The first two restrictions on $h_{n}$ are also standard. The third restriction on $h_{n}$, that $\inf_{u \in [0,1]} n^{\gamma/4}r_{n}(u) \to \infty$, is new. It ensures that the sums used to estimate $\hat{\beta}$ and $\widehat{\lambda(w_{i})}$ diverge with $n$. If $p_{w_i}$ was a continuously distributed $d$-dimensional random vector then, under certain conditions, $P(||p_{w_i}-p_{w_j}||^{2}_{2} \leq h_{n}| \hspace{1mm}w_{i})$ would be on the order of $h_{n}^{d/2}$. The number of agents with codegree function similar to that of agent $i$ would be on the order of $nh_{n}^{d/2}$, and $h_{n}$ could be chosen so that $nh_{n}^{d/2} \to \infty$. Such an assumption, which requires knowledge of the dimension of $p_{w_{i}}$, is standard. Since $p_{w_i}$ is an unknown function, $P(||p_{w_i}-p_{w_j}||_{2}^{2} \leq h_{n}|\hspace{1mm}w_{i})$ can not necessarily be approximated by a polynomial of $h_{n}$ of known order and so $\inf_{u \in [0,1]}n^{\gamma/4}r_{n}(u) \to \infty$ is explicitly assumed instead. One can verify it in practice (in the same sense that one can choose $h_{n}$ to satisfy the first two conditions) by computing $\min_{i=1,...,n}\frac{1}{n}\sum_{j=1}^{n}K\left(\frac{\hat{\delta}^{2}_{ij}}{h_{n}}\right)$ and choosing $h_{n}$ so that it is large relative to $n^{-\gamma/4}$. 



Proposition 2 states that Assumptions 1-4 are sufficient for $\hat{\beta}$ and $\widehat{\lambda(w_{i})}$ to be consistent.
 \begin{flushleft} 
\textbf{Proposition 2}: Under Assumptions 1-4, $\left(\hat{\beta} - \beta\right) \to_{p} 0$ and $\max_{i = 1,...,n}\left|\widehat{\lambda(w_{i})} - \lambda(w_{i})\right| \to_{p} 0$ as $n \to \infty$.
\end{flushleft}

The proof of Proposition 2 is complicated by the fact that codegree functions are not finite dimensional. There is no adequate notion of a density for the distribution of codegree functions, which plays a key role in the standard theory \cite*[see generally][]{ferraty2006nonparametric}. Furthermore, even when the functions $f$ and $\lambda$ are relatively smooth, the bias of $\hat{\beta}$ may still be large relative to its variance. To make reliable inferences about $\beta$ using $\hat{\beta}$, I recommend a bias correction. See  \cite{auerbach2021identification}  for details.

\section{Conclusion}
This paper proposes a new way to incorporate network data into econometric modeling. An unobserved covariate called social influence is determined by an agent's link function, which describes the collection of probabilities that the agent is linked to other agents in the population. Estimation is based on matching pairs of agents with similar columns of the squared adjacency matrix. 

Understanding how to incorporate different kinds of network data into econometric modeling is an important avenue for future research. A contribution of this paper is to demonstrate that in some cases identification and estimation is possible without strong parametric assumptions about how the network is generated or exactly which features of the network determine the outcome of interest.

\bibliographystyle{chicago}
\bibliography{literature}
\appendix
\small
\section{Appendix}

\subsection{Network sparsity}
The network formation model (\ref{eq3}) implies that $D$ is almost surely dense or empty in the limit. That is, for a fixed $f$ and as $n$ tends to infinity, the fraction of realized links in the network converges to $\int\int f(u,v)dudv$ which is either positive or zero.  

Many networks of interest to economists are sparse, however, in the sense that relatively few agent-pairs in the population interact. The framework of this paper can potentially accommodate sparsity by allowing some parameters of the model to vary with the sample size, for instance
\begin{align*}
y_{i} &= x_{i}\beta + \lambda_{n}(w_{i}) + \varepsilon_{i} \\
D_{ij} &= \mathbbm{1}\{\eta_{ij} \leq f_{n}(w_{i},w_{j}))\}\mathbbm{1}\{i \neq j\}
\end{align*} 
where $\lambda_{n}$ and $f_{n}$ now depend on $n$ \citep[see][Section 3.8]{graham2019network}. The fraction of realized links in the network converges to $\int\int f_{n}(u,v)dudv$ which can be arbitrarily small as $n$ grows large. Allowing the agent link functions $f_{n}(w_{i},\cdot) : [0,1] \to [0,1]$ to depend on $n$ does not alter the results of Section 2 in that,  \emph{mutatis mutandis}, Assumptions 1-4 still imply Propositions 1 and 2. 

But while the results of Section 2 may hold under network sparsity, Assumption 3 is potentially violated. This is because the premise of that assumption is that agents with similar link functions have similar social influence. If $\int\int f_{n}(u,v)dudv$  is shrinking to $0$, then the agent link functions are shrinking to the constant $0$ function, and so it implies that relatively small deviations in the agent link functions are sufficient to distinguish agents with different social influences. When this assumption is implausible, alternative assumptions about link formation or better quality data on agent interactions may be necessary.

\subsection{Lemma 1}
\textbf{Proof of Lemma 1}:  Assume $0 \leq \inf_{u,v \in [0,1]}f(u,v) \leq \sup_{u,v \in [0,1]}f(u,v) \leq 1$. The first claim that $||p_{w_{i}} - p_{w_{j}}||_{2} \leq ||f_{w_{i}} - f_{w_{j}}||_{2}$ for every $i, j \in \{1,...,n\}$ is almost an immediate consequence of Jensen's inequality
\begin{align*}
||p_{w_{i}} - p_{w_{j}}||_{2}^{2} &= \int \left( \int f(\tau,s)\left(f(w_{i},s)-f(w_{j},s)\right)ds\right)^{2}d\tau \\
&\leq \int\int \left(f(\tau,s)\left(f(w_{i},s)-f(w_{j},s)\right)\right)^{2}ds d\tau \\
&\leq \int \left(f(w_{i},s)-f(w_{j},s)\right)^{2}ds =||f_{w_{i}} - f_{w_{j}}||_{2}^{2}
\end{align*}
where the first inequality is due to Jensen and the second is because $\sup_{u,v \in [0,1]}f(u,v)^{2} \leq 1$. \\

Now assume $\inf_{u \in [0,1]}\int \mathbbm{1}\left\{v \in [0,1]: \sup_{\tau \in [0,1]}\left|f(u,\tau)-f(v,\tau)\right| < \epsilon\right\}dv > 0$ for every $\epsilon > 0$ and $0 \leq \inf_{u,v \in [0,1]}f(u,v) \leq \sup_{u,v \in [0,1]}f(u,v) \leq 1$. To demonstrate the second claim that for every $i, j \in \{1,...,n\}$ and $\varepsilon > 0$ there exists a $\delta > 0$ such that $||f_{w_{i}}-f_{w_{j}}||_{2}\mathbbm{1}\{||p_{w_{i}}-p_{w_{j}}||_{2} \leq \delta\} \leq \varepsilon$, I show that for every $i, j \in \{1,...,n\}$ and $\varepsilon > 0$ there exists a $\delta > 0$ such that $\mathbbm{1}\{||f_{w_{i}}-f_{w_{j}}||_{2} > \varepsilon\}  \leq \mathbbm{1}\{||p_{w_{i}}-p_{w_{j}}||_{2} > \delta\}$. The claim then follows
\begin{align*}
||f_{w_{i}}-f_{w_{j}}||_{2}\mathbbm{1}\{||p_{w_{i}}-p_{w_{j}}||_{2} \leq \delta\} 
&\leq \varepsilon + ||f_{w_{i}}-f_{w_{j}}||_{2}\mathbbm{1}\{||p_{w_{i}}-p_{w_{j}}||_{2} \leq \delta\}\mathbbm{1}\{||f_{w_{i}}-f_{w_{j}}||_{2} > \varepsilon\}\\
&\leq \varepsilon + ||f_{w_{i}}-f_{w_{j}}||_{2}\mathbbm{1}\{||p_{w_{i}}-p_{w_{j}}||_{2} \leq \delta\}\mathbbm{1}\{||p_{w_{i}}-p_{w_{j}}||_{2} > \delta\} \\
&= \varepsilon.
\end{align*}

Fix $i,j \in \{1,...,n\}$ and $\varepsilon >0$. Define $x := \text{argmax}_{w \in \{w_{i},w_{j}\}}\left| \int f(w,\tau)(f(w_i,\tau)-f(w_j,\tau))d\tau \right|$ and $S(u,\varepsilon') := \{v \in [0,1] : \sup_{\tau \in [0,1]}\left|f(u,\tau)-f(v,\tau)\right| < \varepsilon'\}$ for any $u \in [0,1]$ and $\varepsilon' > 0$. Let $y$ be an arbitrary element of $S(x,\varepsilon^2/4)$. Then
\begin{align*}
\mathbbm{1}\left\{||f_{w_i}-f_{w_j}||_{2}  > \varepsilon\right\} 
&= \mathbbm{1}\left\{\int\left(f(w_i,\tau)-f(w_j,\tau)\right)^{2}d\tau > \varepsilon^{2}\right\} \\
&= \mathbbm{1}\left\{\int f(w_i,\tau)(f(w_i,\tau)-f(w_j,\tau))d\tau - \int f(w_j,\tau)(f(w_{i},\tau)-f(w_j,\tau))d\tau  > \varepsilon^{2}\right\} \\
&\leq \mathbbm{1}\left\{\left| \int f(x,\tau)(f(w_i,\tau)-f(w_j,\tau))d\tau \right| > \varepsilon^{2}/2\right\}\\
&= \mathbbm{1}\left\{\left| \int \left[f(x,\tau) -f(y,\tau) + f(y,\tau)\right](f(w_i,\tau)-f(w_j,\tau))d\tau \right| > \varepsilon^{2}/2\right\} \\ 
&\leq \mathbbm{1}\left\{ \left|\int f(y,\tau)(f(w_i,\tau)-f(w_j,\tau))d\tau\right|  \right.\\
&\left.\hspace{20mm}+ \left|\int \left( f(x,\tau)-f(y,\tau)\right)(f(w_i,\tau)-f(w_j,\tau))d\tau\right| > \varepsilon^{2}/2\right\}  \\
&\leq \mathbbm{1}\left\{\left|\int f(y,\tau)(f(w_i,\tau)-f(w_j,\tau))d\tau \right| > \varepsilon^{2}/4\right\} \\
&= \mathbbm{1}\left\{\left[\int f(y,\tau)(f(w_i,\tau)-f(w_j,\tau))d\tau \right]^{2} > \varepsilon^{4}/16\right\}
\end{align*}
where the second inequality is due to the triangle inequality and the third inequality is because $y \in S(x,\varepsilon^{2}/4)$ and $0 \leq \inf_{u,v \in [0,1]}f(u,v) \leq \sup_{u,v \in [0,1]}f(u,v) \leq 1$ implies that $\sup_{u,v,\tau \in [0,1]}|f(u,\tau)- f(v,\tau)| \leq 1$ and  $\left|\int (f(x,\tau)-f(y,\tau))(f(w_i,\tau)-f(w_j,\tau))d\tau \right| < \varepsilon^{2}/4$. \\

Define $\omega(\varepsilon') := \inf_{u \in [0,1]}\int \mathbbm{1}\left\{ v \in S(u,\varepsilon')\right\}dv$ where $\omega(\varepsilon') > 0$ for every $\varepsilon' > 0$ by assumption. Since the choice of $i, j \in \{1,...,n\}$, $\varepsilon > 0$, and $y\in S(x,\varepsilon^2/4)$ was arbitrary, it follows that 
\begin{align*}
\mathbbm{1}\left\{||f_{w_i}-f_{w_j}||_{2} > \varepsilon\right\} 
&\leq  \mathbbm{1}\left\{\int\left[\int f(y,\tau)(f(w_i,\tau)-f(w_j,\tau))d\tau \right]^{2}\frac{\mathbbm{1}\{y \in S(x,\varepsilon^{2}/4)\}}{\int \mathbbm{1}\{v \in S(x,\varepsilon^{2}/4)\} dv}dy >  \frac{\varepsilon^{4}}{16}\right\} \\
&\leq \mathbbm{1}\left\{\int\left[\int f(y,\tau)(f(w_i,\tau)-f(w_j,\tau))d\tau \right]^{2}\mathbbm{1}\{y \in S(x,\varepsilon^{2}/4)\}dy >  \frac{\varepsilon^{4}}{16}\times \omega(\varepsilon^{2}/4)\right\} \\
&\leq \mathbbm{1}\left\{\int\left[\int f(y,\tau)(f(w_i,\tau)-f(w_j,\tau))d\tau \right]^{2}dy >  \frac{\varepsilon^{4}}{16}\times \omega(\varepsilon^{2}/4)\right\} \\
&\leq \mathbbm{1}\left\{||p_{w_i}-p_{w_j}||_{2} >   \frac{\varepsilon^{2}}{4}\times \sqrt{\omega(\varepsilon^{2}/4)} \right\}.
\end{align*}
for any $i, j \in \{1,...,n\}$ and $\varepsilon > 0$. The claim follows. $\square$ \\

\textbf{Proof of Lemma A1}: Assume $0 \leq \inf_{u,v \in [0,1]}f(u,v) \leq \sup_{u,v \in [0,1]}f(u,v) \leq 1$. The lower bound follows from the first part of Lemma 1. Now assume the existence of an $\alpha, C > 0$ such that $\inf_{u \in [0,1]}\int \mathbbm{1}\left\{v \in [0,1]: \sup_{\tau \in [0,1]}\left|f(u,\tau)-f(v,\tau)\right| < \varepsilon\right\}dv  \geq \left(\frac{\varepsilon}{C}\right)^{1/\alpha}$ for every $\varepsilon \in [0,1]$ and $0 \leq \inf_{u,v \in [0,1]}f(u,v) \leq \sup_{u,v \in [0,1]}f(u,v) \leq 1$. The second claim that for every $i, j \in \{1,...,n\}$
\begin{align*}
||f_{w_{i}} - f_{w_{j}}||_{2} \leq 2C^{\frac{1}{2+4\alpha}} ||p_{w_{i}} - p_{w_{j}}||_{2}^{\frac{\alpha}{1+2\alpha}}
\end{align*}
follows from the second part of Lemma 1 by replacing $\omega(\varepsilon')$ with $\left(\frac{\varepsilon'}{C}\right)^{1/\alpha}$. Specifically, $\omega(\varepsilon^{2}/4) := \inf_{u \in [0,1]}\int \mathbbm{1}\left\{v \in [0,1]: \sup_{\tau \in [0,1]}\left|f(u,\tau)-f(v,\tau)\right| <  \varepsilon^{2}/4\right\}dv  \geq \left(\frac{\varepsilon^{2}}{4C}\right)^{1/\alpha}$. As a result,
\begin{align*}
 \mathbbm{1}\left\{||f_{w_{i}}-f_{w_j}||_{2} > \varepsilon\right\} 
 &\leq \mathbbm{1}\left\{ ||p_{w_i}-p_{w_j}||_{2} >  \frac{\varepsilon^{2}}{4}\left(\frac{\varepsilon^{2}}{4C}\right)^{1/2\alpha}\right\} \\
&= \mathbbm{1}\left\{ 2C^{\frac{1}{2+4\alpha}} ||p_{w_{i}} - p_{w_{j}}||_{2}^{\frac{\alpha}{1+2\alpha}} > \varepsilon \right\}
\end{align*}
for any  $i, j \in \{1,...,n\}$ and $\varepsilon >0$, and so 
\begin{align*}
||f_{w_{i}}-f_{w_j}||_{2} - \eta < 2C^{\frac{1}{2+4\alpha}} ||p_{w_{i}} - p_{w_{j}}||_{2}^{\frac{\alpha}{1+2\alpha}}
\end{align*}
for any  $i, j \in \{1,...,n\}$ and  $\eta > 0$. The claim follows. $\square$

\subsection{Proof of Proposition 1}
\begin{flushleft}
\textbf{Proof of Proposition 1}: Let $d_{ij} := ||f_{w_{i}}-f_{w_{j}}||_{2}$ and $u_{i} := y_{i} - x_{i}\beta = \lambda(w_{i}) + \varepsilon_{i}$. I first demonstrate claim (ii) that $E\left[u_{i}|f_{w_{i}}\right]  = \lambda(w_{i})$. This claim follows 
\begin{align*}
E\left[u_{i}|f_{w_{i}}\right] = E\left[\lambda(w_{i})|f_{w_{i}}\right] + E\left[E\left[\varepsilon_{i}| w_{i}\right]|f_{w_{i}}\right] = \lambda(w_{i})
\end{align*}
where $E\left[E\left[\varepsilon_{i}| w_{i}\right]|f_{w_{i}}\right] = 0$ because $E\left[\varepsilon_{i}|x_{i},w_{i}\right] = 0$ by Assumption 1 and $E\left[\lambda(w_{i})|f_{w_{i}}\right]  = \lambda(w_{i})$ because 
\begin{align*}
\left|E\left[\lambda(w_{i})|f_{w_{i}}\right] - \lambda(w_{i})\right| 
&= \left|E\left[\left(\lambda(w) - \lambda(w_{i})\right)| \hspace{1mm}||f_{w}- f_{w_{i}}||_{2} = 0, w_{i}\right]\right| \\
&\leq \left(E\left[\left(\lambda(w) - \lambda(w_{i})\right)^{2}|\hspace{1mm} ||f_{w}- f_{w_{i}}||_{2} = 0, w_{i}\right]\right)^{1/2} \\
&= \left(E\left[\left(\lambda(w) - \lambda(w_{i})\right)^{2}\mathbbm{1}\{ ||f_{w}- f_{w_{i}}||_{2} \leq \delta\}| \hspace{1mm} ||f_{w}- f_{w_{i}}||_{2} = 0, w_{i}\right]\right)^{1/2} \text{ for every } \delta > 0\\
&\leq \epsilon^{1/2} \text{ for every } \epsilon > 0
\end{align*}
where $w$ is an independent copy of $w_{i}$, the first equality is due to the definition of $E\left[\lambda(w_{i})|f_{w_{i}}\right]$, the first inequality is due to Jensen, and the last inequality is due to Assumption 3. \newline \\


I now demonstrate claim (i) that $\beta$ uniquely minimizes $E\left[\left(y_{i}-y_{j} - (x_{i}-x_{j})b\right)^{2}|\hspace{1mm}d_{ij} = 0\right]$ over $b \in \mathbb{R}^{k}$. This claim follows from expanding the square
\begin{align*}
&E\left[\left(y_{i}-y_{j} - (x_{i}-x_{j})b\right)^{2}|d_{ij} = 0\right] = E\left[\left((x_{i}-x_{j})(\beta-b) + (u_{i}-u_{j})\right)^{2}|d_{ij} = 0\right]\\
&= (\beta-b)'E[(x_{i}-x_{j})'(x_{i}-x_{j}) |d_{ij} = 0](\beta - b) + E[(u_{i}-u_{j})^{2}|d_{ij} = 0] \\
&\hspace{20mm} +2(\beta - b)'E[(x_{i}-x_{j})'(u_{i}-u_{j}) |d_{ij} = 0].
\end{align*}
The first summand is uniquely minimized at $b = \beta$ by Assumption 2 (see below), the second summand does not depend on $b$, and the third summand is equal to $0$ for any $b \in \mathbb{R}^{k}$ since $E\left[\varepsilon_{i}|x_{i},w_{i}\right] = 0$ by Assumption 1 and $2(\beta - b)'E[(x_{i}-x_{j})'(\lambda(w_{i})-\lambda(w_{j})) |d_{ij} = 0] = 0$ by Assumption 3 following the same logic as in the proof of claim (ii)
\begin{align*}
&\left|(\beta - b)'E\left[(x_{i}-x_{j})'(\lambda(w_{i})-\lambda(w_{j}))|d_{ij} = 0\right]\right| \\
&\hspace{20mm}\leq |(\beta-b)'E\left[(x_{i}-x_{j})'(x_{i}-x_{j})|d_{ij} = 0\right](\beta-b)|^{1/2}E\left[(\lambda(w_{i})-\lambda(w_{j}))^{2}|d_{ij} = 0\right]^{1/2}\\
&\hspace{20mm}= |(\beta-b)'E\left[(x_{i}-x_{j})'(x_{i}-x_{j})|d_{ij} = 0\right](\beta-b)|^{1/2}\\
&\hspace{40mm}\times E\left[(\lambda(w_{i})-\lambda(w_{j}))^{2}\mathbbm{1}\{d_{ij} \leq \delta\}|d_{ij} = 0\right]^{1/2} \text{ for every } \delta > 0\\
&\hspace{20mm}\leq \epsilon^{1/2} \text{ for every } \epsilon > 0
\end{align*}
where the first inequality is due to Cauchy-Schwarz and the last inequality is due to Assumption 3. \newline \\

To see that Assumption 2 implies that $ (\beta-b)'E[(x_{i}-x_{j})'(x_{i}-x_{j}) |d_{ij} = 0](\beta - b)$ is uniquely minimized at $b = \beta$, write $\mu_{i} = E\left[x_{i}|w_{i}\right]$,  $\xi_{i} = x_{i} - \mu_{i}$, and
\begin{align*}
E[(x_{i}-x_{j})'(x_{i}-x_{j}) | d_{ij} = 0] &= E[(\mu_{i}-\mu_{j})'(\mu_{i}-\mu_{j}) | d_{ij} = 0] + E[(\xi_{i}-\xi_{j})'(\xi_{i}-\xi_{j}) | d_{ij} = 0] \\
&= E[(\mu_{i}-\mu_{j})'(\mu_{i}-\mu_{j}) | d_{ij} = 0] + E\left[E\left[\xi_{i}'\xi_{i}|w_{i}\right] + E\left[\xi_{j}'\xi_{j}|w_{j}\right]  | d_{ij} = 0\right]
\end{align*}
where both equalities are due to the fact that $E\left[\xi_{i}|x_{i},w_{i}\right] = 0$. The first summand $E[(\mu_{i}-\mu_{j})'(\mu_{i}-\mu_{j}) | d_{ij} = 0]$ is positive semidefinite. The second summand $E\left[E\left[\xi_{i}'\xi_{i}|w_{i}\right] + E\left[\xi_{j}'\xi_{j}|w_{j}\right]  | d_{ij} = 0\right]$ is positive definite by Assumption 2. It follows that $E[(x_{i}-x_{j})'(x_{i}-x_{j}) | d_{ij} = 0]$ is positive definite and so $(\beta-b)'E[(x_{i}-x_{j})'(x_{i}-x_{j}) |d_{ij} = 0](\beta - b)$ is nonnegative for all $b \in \mathbb{R}^{k}$ and zero only when $b = \beta$. $\square$

\end{flushleft}
%

%
%

\subsection{Proof of Proposition 2}
The proof of Proposition 2 relies on the following Lemma B1. 
\begin{flushleft}
\textbf{Lemma B1}: Suppose Assumptions 1 and 4. Then
\begin{align*}
\max_{i \neq j}\left| \hat{\delta}^{2}_{ij} - ||p_{w_{i}}- p_{w_{j}}||^{2}_{2} \right| = o_{p}\left(n^{-\gamma/4}h_{n}\right).
\end{align*}
where $\gamma > 0$ is the constant from Assumption 4. 
\end{flushleft}

\begin{flushleft}
\textbf{Proof of Lemma B1}:
Let $h_{n}' := n^{-\gamma/4}h_{n}$, $p_{w_iw_j} := p(w_{i},w_{j}) = \int f_{w_{i}}(\tau)f_{w_{j}}(\tau)d\tau$, $p_{w_{i}}(s) := p(w_{i},s)$, $\hat{p}_{w_{i}w_{j}} := \frac{1}{n}\sum_{t=1}^{n}D_{it}D_{jt}$, $||\hat{p}_{w_{i}} - p_{w_{i}}||^{2}_{2,n} := \frac{1}{n}\sum_{s=1}^{n}\left(\hat{p}_{w_{i}w_{s}} - p_{w_{i}w_{s}}\right)^{2}$, and $||p_{w_{i}}-p_{w_{j}}||^{2}_{2,n} := \frac{1}{n}\sum_{s=1}^{n}\left(p_{w_{i}w_{s}} - p_{w_{j}w_{s}}\right)^{2}$. Then for any fixed $\epsilon > 0$
\begin{align*}
&P\left(\max_{i \neq j}h_{n}'^{-1} \left| \hat{\delta}_{ij}^{2} - ||p_{w_{i}}-p_{w_{j}}||^{2}_{2}  \right| > \epsilon \right) \\
&= P\left(\max_{i \neq j} h_{n}'^{-1} \left| \hat{\delta}_{ij}^{2} - ||p_{w_{i}}-p_{w_{j}}||^{2}_{2,n} + ||p_{w_{i}}-p_{w_{j}}||^{2}_{2,n} - ||p_{w_{i}}-p_{w_{j}}||^{2}_{2}  \right| > \epsilon \right) \\
&\leq P\left(\max_{i \neq j}h_{n}'^{-1} \left| \hat{\delta}^{2}_{ij} - ||p_{w_{i}}-p_{w_{j}}||^{2}_{2,n} \right|  > \epsilon/2 \right)  \\
&\hspace{30 mm} + P\left(\max_{i \neq j} h_{n}'^{-1}\left| ||p_{w_{i}}-p_{w_{j}}||^{2}_{2,n} - ||p_{w_{i}}-p_{w_{j}}||^{2}_{2}  \right|  > \epsilon/2 \right) \\
&= P\left(\max_{i \neq j}h_{n}'^{-1} \left| \hat{\delta}^{2}_{ij} - ||p_{w_{i}}-p_{w_{j}}||^{2}_{2,n} \right|  > \epsilon/2 \right) + o(1)  \\
&\leq P\left(\max_{i \neq j} h_{n}'^{-1}\frac{1}{n}\sum_{s=1}^{n}\left|(\hat{p}_{w_{i}w_{s}}-\hat{p}_{w_{j}w_{s}}) -(p_{w_{i}w_{s}}-p_{w_{j}w_{s}})\right|  > \epsilon/8 \right) + o(1) \\
&\leq 2P\left(\max_{i} h_{n}'^{-1}\frac{1}{n}\sum_{s=1}^{n}\left|\hat{p}_{w_{i}w_{s}} - p_{w_{i}w_{s}}\right|  > \epsilon/16 \right) + o(1) \\
&= o(1)
\end{align*} 
where $P\left(\max_{i \neq j} h_{n}'^{-1}\left| ||p_{w_{i}}-p_{w_{j}}||^{2}_{2,n} - ||p_{w_{i}}-p_{w_{j}}||^{2}_{2}  \right|  > \epsilon/2 \right) = o(1)$ in the second equality and $P\left(\max_{i} h_{n}'^{-1}\frac{1}{n}\sum_{s=1}^{n}\left|\hat{p}_{w_{i}w_{s}} - p_{w_{i}w_{s}}\right|  > \epsilon/16 \right) = o(1)$ in the third equality are demonstrated below, the first inequality is due to the triangle inequality and the union bound, the second inequality is due Jensen and the fact that $\sup_{u\in[0,1]}\left|p_{u} + \hat{p}_{u}\right| \leq 2$, and the third inequality is due to the triangle inequality. 
\newline

The third equality, that $P\left(\max_{i} h_{n}'^{-1}\frac{1}{n}\sum_{s=1}^{n}\left|\hat{p}_{w_{i}w_{s}} - p_{w_{i}w_{s}}\right|  > \epsilon/16 \right) = o(1)$ follows from the fact that $\max_{i \neq j}h_{n}'^{-1}|\hat{p}_{w_{i}w_{j}}-p_{w_{i}w_{j}}| \to_{p} 0$ by Bernstein's inequality and the union bound. Specifically, Bernstein's inequality implies that for any $\epsilon > 0$
\begin{align*}
P\left(h_{n}'^{-1}|\hat{p}_{w_{i}w_{j}}-p_{w_{i}w_{j}}| > \epsilon\right) &= P\left(h_{n}'^{-1}\left|\frac{1}{n}\sum_{t=1}^{n}\left(D_{it}D_{jt} - p_{w_{i}w_{j}}\right) \right| > \epsilon \right)\\ 
&\leq P\left(h_{n}'^{-1}\frac{n-2}{n}\left|\frac{1}{n-2}\sum_{t\neq i,j}\left(D_{it}D_{jt} - p_{w_{i}w_{j}}\right) - \frac{2}{n-2}p_{w_{i}w_{j}} \right| > \epsilon \right) \\
&\leq 2\exp \left(\frac{-(n-2)(h_n'\frac{n}{n-2}\epsilon - \frac{2}{n-2})^{2}}{2 + \frac{2}{3}\left(h_{n}'\frac{n}{n-2}\epsilon - \frac{2}{n-2}\right)}\right)
\end{align*}
and the union bound gives 
\begin{align*}
P\left(\max_{i \neq j}h_{n}'^{-1}|\hat{p}_{w_{i}w_{j}}-p_{w_{i}w_{j}}| > \epsilon\right) \leq 2n(n-1) \exp \left(\frac{-(n-2)(h_n'\frac{n}{n-2}\epsilon - \frac{2}{n-2})^{2}}{2 + \frac{2}{3}\left(h_{n}'\frac{n}{n-2}\epsilon - \frac{2}{n-2}\right)}\right)
\end{align*}
which is $o\left(n^2\exp\left(-n^{\gamma/2}\right)\right)$ and so $o(1)$ since $h_{n}' \to 0$ and $n^{1-\gamma/2}h_{n}'^{2}\to \infty$ by Assumption 4.
\newline 

The second equality, that $P\left(\max_{i \neq j} h_{n}'^{-1}\left| ||p_{w_{i}}-p_{w_{j}}||^{2}_{2,n} - ||p_{w_{i}}-p_{w_{j}}||^{2}_{2}  \right|  > \epsilon/2 \right) = o(1)$, also follows from Bernstein's inequality and the union bound because
\begin{align*}
&P\left(h_{n}'^{-1}\left| ||p_{w_{i}}-p_{w_{j}}||^{2}_{2,n} - ||p_{w_{i}}-p_{w_{j}}||^{2}_{2} \right| > \epsilon \right) \\
& = P\left(h_{n}'^{-1}\left|\frac{1}{n}\sum_{s=1}^{n}\left(p_{w_{i}w_{s}} - p_{w_{j}w_{s}}\right)^{2} - \int \left(p_{w_{i}}(s)-p_{w_{j}}(s)\right)^{2}ds\right| > \epsilon \right) \\
&\leq 2\exp\left(\frac{-n\left(h_{n}'\epsilon\right)^{2}}{2 + \frac{2}{3}h_{n}'\epsilon} \right) 
\end{align*}
which is $o(1)$ since $h_{n}' \to 0$ and $nh_{n}'^{2} \to \infty$ by Assumption 4. This completes the proof. $\square$ \newline
\end{flushleft}

\begin{flushleft}
\textbf{Proof of Proposition 2}: I start with the first result that $\left(\hat{\beta} - \beta\right) \to_{p} 0$. Let $u_{i} := y_{i} - x_{i}\beta = \lambda(w_{i}) + \varepsilon_{i}$, $\delta_{ij} := \delta(w_{i},w_{j}) = ||p_{w_{i}}-p_{w_{j}}||_{2}$, $r_{n} := \int r_{n}(u)du = E\left[K\left(\frac{\delta^{2}_{ij}}{h_{n}}\right)\right]$, $\Gamma_{n} := r_{n}^{-1}E\left[(x_{i}-x_{j})'(x_{i}-x_{j})K\left(\frac{\delta^{2}_{ij}}{h_{n}}\right)\right]$, and write
\begin{align*}
\hat{\beta} = \beta + &\left( \sum_{i=1}^{n-1}\sum_{j =  i+1}^{n}(x_{i}-x_{j})'(x_{i}-x_{j})K\left(\frac{\hat{\delta}^{2}_{ij}}{h_{n}}\right)\right)^{-1}\left(\sum_{i=1}^{n-1}\sum_{j = i + 1}^{n}(x_{i}-x_{j})'(u_{i}-u_{j})K\left(\frac{\hat{\delta}^{2}_{ij}}{h_{n}}\right)\right).
\end{align*}
I first show that $\left|\left({n \choose 2}r_{n}\right)^{-1}\sum_{i=1}^{n-1}\sum_{j =  i+1}^{n}(x_{i}-x_{j})'(x_{i}-x_{j})K\left(\frac{\hat{\delta}^{2}_{ij}}{h_{n}}\right) - \Gamma_{n}\right| \to_{p} 0$ where $r_{n} > 0$ by Assumption 4. Nearly identical arguments yield  $\left|\left({n \choose 2}r_{n}\right)^{-1}\sum_{i=1}^{n-1}\sum_{j =  i+1}^{n}(x_{i}-x_{j})'(u_{i}-u_{j})K\left(\frac{\hat{\delta}^{2}_{ij}}{h_{n}}\right)\right| \to_{p} 0$, and so the claim follows since Assumption 2 implies that the eigenvalues of $\Gamma_{n}$ are bounded away from $0$ (see below).
\newline

Let $D_{n} := \left({n \choose 2}r_{n}\right)^{-1}\sum_{i}\sum_{j}(x_{i}-x_{j})'(x_{i}-x_{j})K\left(\frac{\hat{\delta}^{2}_{ij}}{h_{n}}\right)$. By the mean value theorem and smoothness condition on the kernel function in Assumption 4, $D_{n} = \left({n \choose 2}r_{n}\right)^{-1}\sum_{i}\sum_{j}(x_{i}-x_{j})'(x_{i}-x_{j})\left[K\left(\frac{\delta_{ij}^{2}}{h_{n}}\right) + K'\left(\frac{\iota_{ij}}{h_{n}}\right)\left(\frac{\hat{\delta}^{2}_{ij}-\delta^{2}_{ij}}{h_{n}} \right) \right]$
where $\{\iota_{ij}\}_{i \neq j}$ are the mean values implied by that theorem. By Lemma B1, $\max_{i \neq j} \frac{\hat{\delta}^{2}_{ij}-\delta^{2}_{ij}}{h_{n}} = o_{p}\left(n^{-\gamma/4} \right)$ and so  $\left({n \choose 2}r_{n}\right)^{-1}\sum_{i}\sum_{j}(x_{i}-x_{j})'(x_{i}-x_{j})K'\left(\frac{\iota_{ij}}{h_{n}}\right)\left(\frac{\hat{\delta}^{2}_{ij}-\delta^{2}_{ij}}{h_{n}} \right) = o_{p}(1)$ because $K'$ is absolutely bounded, $x_{i}$ has finite second moments, and $r_{n}n^{\gamma/4} \to \infty$ by Assumption 4. It follows that $D_{n} = \left({n \choose 2}r_{n}\right)^{-1}\sum_{i}\sum_{j}(x_{i}-x_{j})'(x_{i}-x_{j})K\left(\frac{\delta^{2}_{ij}}{h_{n}}\right) + o_{p}(1)$.
\newline

Let $D_{n}' := \left({n \choose 2}r_{n}\right)^{-1}\sum_{i}\sum_{j}(x_{i}-x_{j})'(x_{i}-x_{j})K\left(\frac{\delta^{2}_{ij}}{h_{n}}\right)$. $D_{n}'$ is a second order U-statistic with kernel depending on $n$, in the sense of \cite{ahn1993semiparametric}. Their Lemma A.3 (i) implies that $D_{n}' = r_{n}^{-1}E\left[(x_{i}-x_{j})'(x_{i}-x_{j})K\left(\frac{\delta^{2}_{ij}}{h_{n}}\right)\right] + o_{p}(1)$
since $nr_{n}^{2} \to \infty$ by Assumption 4. So $D_{n} = \Gamma_{n} + o_{p}(1)$.
\newline

Let  $U_{n} := \left({n \choose 2}r_{n}\right)^{-1}\sum_{i}\sum_{j}(x_{i}-x_{j})'(u_{i}-u_{j})K\left(\frac{\hat{\delta}^{2}_{ij}}{h_{n}}\right)$. A nearly identical argument gives  $U_{n} = r_{n}^{-1}E\left[(x_{i}-x_{j})'(u_{i}-u_{j})K\left(\frac{\delta^{2}_{ij}}{h_{n}}\right)\right]  + o_{p}(1)$. Furthermore, 
\begin{align*}
r_{n}^{-1}E\left[(x_{i}-x_{j})'(u_{i}-u_{j})K\left(\frac{\delta^{2}_{ij}}{h_{n}}\right)\right] 
&= r_{n}^{-1}E\left[(x_{i}-x_{j})'(\lambda(w_{i})-\lambda(w_{j}))K\left(\frac{\delta^{2}_{ij}}{h_{n}}\right)\right] \\
&= r_{n}^{-1}E\left[(x_{i}-x_{j})'(\lambda(w_{i})-\lambda(w_{j}))\mathbbm{1}\{\delta_{ij} \leq h_{n}\}K\left(\frac{\delta^{2}_{ij}}{h_{n}}\right)\right] \\
&= o_{p}(1)
\end{align*}
 where the first equality is because $E\left[\varepsilon_{i}|x_{i},w_{i}\right] = 0$ and the last equality is by Assumptions 3-4, Lemma 1, and because $\sup_{u \in [0,1]}\left|E\left[x_{i}|w_{i}= u\right]\right| < \infty$.
\newline

The result $\left(\hat{\beta} - \beta\right) := D_{n}^{-1}U_{n} = \left(\Gamma_{n}+ \left(D_{n} - \Gamma_{n}\right)\right)^{-1}U_{n} = (\Gamma_{n} + o_{p}(1))^{-1}o_{p}(1) = o_{p}(1)$ follows because the eigenvalues of $\Gamma_{n}$ are bounded away from $0$ (and so $||(\Gamma_{n} + o_{p}(1))^{-1}||$ is bounded). To see this, let  $\sigma_{k}\left(\Gamma_{n}\right)$ denote the smallest eigenvalue of $\Gamma_{n}$, $\mu_{i} := E\left[x_{i}|w_{i}\right]$, and $\xi_{i} := x_{i} - \mu_{i}$. Then   
\begin{align*}
\liminf_{n\to\infty} \sigma_{k}\left(\Gamma_{n}\right) &:= \liminf_{n\to\infty}\sigma_{k}\left(r_{n}^{-1}E\left[(x_{i}-x_{j})'(x_{i}-x_{j})K\left(\frac{\delta_{ij}}{h_{n}}\right)\right]\right)\\
&\geq \liminf_{n\to\infty} r_{n}^{-1}E\left[\sigma_{k}(E\left[(x_{i}-x_{j})'(x_{i}-x_{j})|w_{i},w_{j}\right])K\left(\frac{\delta_{ij}}{h_{n}}\right)\right] \\
&= \liminf_{n\to\infty} r_{n}^{-1}E\left[\sigma_{k}(E\left[(\mu_{i}-\mu_{j})'(\mu_{i}-\mu_{j}) + \xi_{i}'\xi_{i} + \xi_{j}'\xi_{j}|w_{i},w_{j}\right])K\left(\frac{\delta_{ij}}{h_{n}}\right)\right] \\
&\geq \liminf_{n\to\infty} r_{n}^{-1}E\left[\left(\sigma_{k}((\mu_{i}-\mu_{j})'(\mu_{i}-\mu_{j})) + \sigma_{k}(\Xi(w_{i})) + \sigma_{k}(\Xi(w_{j})\right))K\left(\frac{\delta_{ij}}{h_{n}}\right)\right] >0
\end{align*}
where the first and second inequalities are due to Jensen, the second equality is because $E\left[\xi_{i}|w_{i}\right] = 0$, and the last inequality is by Assumption 2 and the fact that $(\mu_{i}-\mu_{j})'(\mu_{i}-\mu_{j})$ is positive semidefinite. 
\newline

I now demonstrate that $\max_{i = 1,..,n}\left|\widehat{\lambda(w_{i})} - \lambda(w_{i})\right| \to_{p} 0$. Let $\delta_{it}$ and $r_{n}(u)$ shorthand $\delta(w_{i},w_{t})$, and $E\left[K\left(\frac{\delta_{it}^{2}}{h_{n}}\right)| w_{i} = u\right]$ respectively. Write
\begin{align*}
\widehat{\lambda(w_{i})} - \lambda(w_{i}) 
&= \frac{\sum_{t=1}^{n}\left(y_{t} - x_{t}\beta - \lambda(w_{i})\right)K\left(\frac{\hat{\delta}^{2}_{it}}{h_{n}}\right)}{\sum_{t=1}^{n}K\left(\frac{\hat{\delta}^{2}_{it}}{h_{n}}\right)}
-  \frac{\sum_{t=1}^{n}x_{t}K\left(\frac{\hat{\delta}^{2}_{it}}{h_{n}}\right)}{\sum_{t=1}^{n}K\left(\frac{\hat{\delta}^{2}_{it}}{h_{n}}\right)}\left(\hat{\beta} - \beta\right) \\
&= \frac{\sum_{t=1}^{n}\left(\varepsilon_{t} + \lambda(w_{t}) - \lambda(w_{i})\right)K\left(\frac{\hat{\delta}^{2}_{it}}{h_{n}}\right)}{\sum_{t=1}^{n}K\left(\frac{\hat{\delta}^{2}_{it}}{h_{n}}\right)}
-  \frac{\sum_{t=1}^{n}x_{t}K\left(\frac{\hat{\delta}^{2}_{it}}{h_{n}}\right)}{\sum_{t=1}^{n}K\left(\frac{\hat{\delta}^{2}_{it}}{h_{n}}\right)}\left(\hat{\beta} - \beta\right).
\end{align*}
\newline

I first consider the denominator $\sum_{t=1}^{n}K\left(\frac{\hat{\delta}^{2}_{it}}{h_{n}}\right)$. Following previous arguments, Lemma B1 and the smoothness conditions on the kernel function in Assumption 4 imply that
\begin{align*}
\max_{i = 1,...,n} \left|\frac{1}{n}\sum_{t=1}^{n}K\left(\frac{\hat{\delta}^{2}_{it}}{h_{n}}\right) - \frac{1}{n}\sum_{t=1}^{n}K\left(\frac{\delta^{2}_{it}}{h_{n}}\right)\right| = o_{p}\left(n^{-\gamma/4}\right)
\end{align*}
while Hoeffding's inequality and the union bound give
\begin{align*}
\max_{i=1,...,n}\left| \frac{1}{n}\sum_{t=1}^{n}K\left(\frac{\delta^{2}_{it}}{h_{n}}\right) - E\left[K\left(\frac{\delta^{2}_{it}}{h_{n}}\right)|w_{i}\right]\right| = o_{p}\left(n^{-1/4}\right).
\end{align*}
It follows from the triangle inequality that
\begin{align*}
\max_{i=1,...,n}\left| \frac{1}{n}\sum_{t=1}^{n}K\left(\frac{\hat{\delta}^{2}_{it}}{h_{n}}\right) - E\left[K\left(\frac{\delta^{2}_{it}}{h_{n}}\right)|w_{i}\right]\right| = o_{p}\left(n^{-\gamma/4}\right) + o_{p}\left(n^{-1/4}\right) = o_{p}\left(n^{-\gamma/4}\right)
\end{align*} 
since the restrictions on the bandwidth in Assumption 4 imply that $\gamma < 1$.
\newline

A nearly identical argument applied to the numerators $\sum_{t=1}^{n}\left(\varepsilon_{t} + \lambda(w_{t}) - \lambda(w_{i})\right)K\left(\frac{\hat{\delta}^{2}_{it}}{h_{n}}\right)$ and $\sum_{t=1}^{n}x_{t}K\left(\frac{\hat{\delta}^{2}_{it}}{h_{n}}\right)$ gives
\begin{align*}
\max_{i=1,...,n}\left| \frac{1}{n}\sum_{t=1}^{n}\left(\varepsilon_{t} + \lambda(w_{t}) - \lambda(w_{i})\right)K\left(\frac{\hat{\delta}^{2}_{it}}{h_{n}}\right) - E\left[\left(\lambda(w_{t}) - \lambda(w_{i})\right)K\left(\frac{\delta^{2}_{it}}{h_{n}}\right)|w_{i}\right]\right| = o_{p}\left(n^{-\gamma/4}\right).
\end{align*} 
and 
 \begin{align*}
\max_{i=1,...,n}\left| \frac{1}{n}\sum_{t=1}^{n}x_{t}K\left(\frac{\hat{\delta}^{2}_{it}}{h_{n}}\right) - E\left[x_{t}K\left(\frac{\delta^{2}_{it}}{h_{n}}\right)|w_{i}\right]\right| = o_{p}\left(n^{-\gamma/4}\right)
\end{align*} 
since $\varepsilon_{t}$ and $x_{t}$ have finite eighth moments, $\sup_{u\in [0,1]}|\lambda(u)| < \infty$, and $E\left[\varepsilon_{t}|x_{t},w_{t}\right] = 0$ by Assumption 1.  
\newline 

It follows that 
\begin{align*}
\max_{i = 1,..,n}\left|\widehat{\lambda(w_{i})} - \lambda(w_{i})\right|
&= \max_{i = 1,..,n}\left| \frac{\sum_{t=1}^{n}\left(\varepsilon_{t} + \lambda(w_{t}) - \lambda(w_{i})\right)K\left(\frac{\hat{\delta}^{2}_{it}}{h_{n}}\right)}{\sum_{t=1}^{n}K\left(\frac{\hat{\delta}^{2}_{it}}{h_{n}}\right)}
-  \frac{\sum_{t=1}^{n}x_{t}K\left(\frac{\hat{\delta}^{2}_{it}}{h_{n}}\right)}{\sum_{t=1}^{n}K\left(\frac{\hat{\delta}^{2}_{it}}{h_{n}}\right)}\left(\hat{\beta} - \beta\right)\right| \\
&\leq  \max_{i = 1,..,n}\left|\frac{\sum_{t=1}^{n}\left(\varepsilon_{t} + \lambda(w_{t}) - \lambda(w_{i})\right)K\left(\frac{\hat{\delta}^{2}_{it}}{h_{n}}\right)}{\sum_{t=1}^{n}K\left(\frac{\hat{\delta}^{2}_{it}}{h_{n}}\right)}\right| 
+ \max_{i=1,...,n} \left| \frac{\sum_{t=1}^{n}x_{t}K\left(\frac{\hat{\delta}^{2}_{it}}{h_{n}}\right)}{\sum_{t=1}^{n}K\left(\frac{\hat{\delta}^{2}_{it}}{h_{n}}\right)}\right|\left|\hat{\beta} - \beta\right|  \\
&= \max_{i = 1,..,n}\left|\frac{E\left[\left(\lambda(w_{t}) - \lambda(w_{i})\right)K\left(\frac{\delta^{2}_{it}}{h_{n}}\right)|w_{i}\right]}{E\left[K\left(\frac{\delta^{2}_{it}}{h_{n}}\right)|w_{i}\right]}\right| 
+ \max_{i=1,...,n} \left| \frac{E\left[x_{t}K\left(\frac{\delta^{2}_{it}}{h_{n}}\right)|w_{i}\right]}{E\left[K\left(\frac{\delta^{2}_{it}}{h_{n}}\right)|w_{i}\right]}\right|\left|\hat{\beta} - \beta\right| \\
&\hspace{20mm} + o_{p}\left(\left(n^{\gamma/4}\inf_{u \in [0,1]}r_{n}(u)\right)^{-1}\right) \\
&= o_{p}(1)
\end{align*}
where, following previous arguments, the term $\max_{i = 1,..,n}\left|\frac{E\left[\left(\lambda(w_{t}) - \lambda(w_{i})\right)K\left(\frac{\delta^{2}_{it}}{h_{n}}\right)|w_{i}\right]}{E\left[K\left(\frac{\delta^{2}_{it}}{h_{n}}\right)|w_{i}\right]}\right| = o_{p}(1)$ by Assumptions 3-4 and Lemma 1, the term $\max_{i=1,...,n} \left| \frac{E\left[x_{t}K\left(\frac{\delta^{2}_{it}}{h_{n}}\right)|w_{i}\right]}{E\left[K\left(\frac{\delta^{2}_{it}}{h_{n}}\right)|w_{i}\right]}\right|\left|\hat{\beta} - \beta\right| = o_{p}(1)$ because  $\sup_{u \in [0,1]}\left|E\left[x_{i}|w_{i}= u\right]\right| < \infty$ and $\left|\hat{\beta} - \beta\right| = o_{p}(1)$, and $n^{\gamma/4}\inf_{u \in [0,1]}r_{n}(u) \to \infty$ by the restrictions on the bandwidth and kernel function in Assumption 4. $\square$

\end{flushleft}

\end{document}